# Nanobob: A Cubesat Mission Concept For Quantum Communication Experiments In An Uplink Configuration

*10 November 2017*

Erik Kerstel [*,1,2], Arnaud Gardelein [3], Mathieu Barthelemy [1,4], The CSUG Team [#,2], Matthias Fink [5], Siddarth Koduru Joshi [5], Rupert Ursin [5,6]

1. Univ. Grenoble Alpes, CNRS, LIPhy, 38000 Grenoble, France
2. Centre Spatial Universitaire de Grenoble, 38000 Grenoble, France
3. Air Liquide Advanced Technologies, Grenoble, France
4. Univ. Grenoble Alpes, IPAG, 38000 Grenoble, France
5. Institute for Quantum Optics and Quantum Information (IQOQI), Austrian Academy of Sciences, Vienna, Austria
6. Vienna Center for Quantum Science and Technology (VCQ), Vienna, Austria

*) erik.kerstel@univ-grenoble-alpes.fr

**KEYWORDS:** CubeSat; satellite; uplink; QKD; quantum; entanglement; cryptography

## SUMMARY

We present a ground-to-space quantum key distribution (QKD) mission concept and the accompanying feasibility study for the development of the low earth orbit CubeSat payload. The quantum information is carried by single photons with the binary codes represented by polarization states of the photons. Distribution of entangled photons between the ground and the satellite can be used to certify the quantum nature of the link: a guarantee that no eavesdropping can take place. By placing the entangled photon source on the ground, the space segments contains "only" the less complex detection system, enabling its implementation in a compact enclosure, compatible with the 12U CubeSat standard (12 dm$^3$). This reduces the overall cost of the project, making it an ideal choice as a pathfinder for future European quantum communication satellite missions. The space segment is also more versatile than one that contains the source since it is compatible with a multiple of QKD protocols (not restricted to entangled photon schemes) and can be used in quantum physics experiments, such as the investigation of entanglement decoherence. Other possible experiments include atmospheric transmission/turbulence characterization, dark area mapping, fine pointing and tracking, and accurate clock synchronization; all crucial for future global scale quantum communication efforts.

---

# The CSUG NanoBob Team is composed of the following engineers, students, and educators who all contributed at different stages to the current study: Yves Gilot (STMicroelectronics), Etienne LeCoarer (UGA), Juana Rodrigo (Rolls Royce), Thierry Sequies (UGA), Vincent Borne (UGA), Guillaume Bourdarot (UGA), Jean-Yves Burlet (UGA), Alexis Christidis (UGA), Jesus Segura (UGA), Benoit Boulanger (UGA), Veronique Boutou (UGA), Mylene Bouzat (Air Liquide), Mathieu Chabanol (UGA), Laurent Fesquet (UGA), Hassen Fourati (UGA), Michel Moulin, Jean-Michel Niot (Air Liquide), Rodrigo Possamai Bastos (UGA), Bogdan Robu (UGA), Etienne Rolland (UGA), and Sylvain Toru (UGA).

## 1.1. INTRODUCTION

Quantum communication is reaching a level of maturity that makes it a practically certain choice for future secure cryptography. In fact, Quantum Key Distribution (QKD) provides a level of communication security that cannot be obtained by classical cryptographic means, including those based on numerical algorithms. The quantum information can be coded into the polarization state of a single photon. The linearity of quantum mechanics leads to the no-cloning theorem, which states that an arbitrary unknown quantum state cannot be copied perfectly [1]. In a properly designed experiment, an eavesdropping attempt by a third party (commonly called "Eve" in the language of cryptography), would necessarily lead to detectable errors. Given our ever-growing reliance on secure data communication, the intrinsic security of quantum communication largely outweighs the disadvantages of additional complexity and cost.

QKD has already been demonstrated to be a practical way to distribute secret keys between two parties in a number of fiber networks, some of them even using existing telecommunication infrastructure (see, e.g., [2] and references therein). However, even in ultra-low loss fibers, losses limit the maximum distance between two parties to a few hundred kilometers, since the no-cloning theorem prohibits the use of simple optical amplifiers. Much progress has been made in the development of quantum repeaters using entanglement swapping over subsections of the overall distance. This requires heralding of successful entanglement creation over the intermediate distances, as well as storage of the entanglement until entanglement has been established in the adjacent link [3]. Taken together this means that quantum repeaters remain a technologically extremely challenging solution.

The alternative of Earth-bound free-space optical links is ultimately limited by Earth's curvature. A more promising approach is to use a satellite terminal that can potentially provide global scale quantum key distribution.

From a more fundamental physics point of view, the same space segment that is the subject of this paper could be used for the investigation of the interaction between entangled photons and the gravitational field [4]. The implied interrelation between Einstein's theory of relativity and Quantum Mechanics presents one of the most interesting questions in modern physics.

Several schemes exist to implement QKD between two parties, a sender named "Alice" and a receiver known as "Bob". The first and probably best-known protocol is due to Bennett and Brassard ("BB84") that proposed a scheme of exchanging a secure key by encoding the quantum information in the polarization state of single photons [5]. In 1991 Arthur Ekert proposed an entanglement-based protocol ("E91") [6], which has the advantage that a simple statistical test ("Bell test") allows one to certify the quantum nature of the link, and therewith its inherent security. Even if Eve controls the source she still cannot obtain information about the key exchanged by Alice and Bob [6, 7, 8]. In E91, of each pair, one photon is directed towards the (local) polarization analyzer and detection module of Alice, the other is directed towards Bob, who just like Alice measures the polarization state of every photon in a randomly chosen basis for each photon and notes its arrival time. In our implementation, both the Alice and Bob detection modules use a 50/50 beam splitter to send the photons randomly to one of two sets of two detectors that define two mutually unbiased bases [9] (identified as horizontal–vertical and diagonal–anti-diagonal, {HV} and {DA}). Alice and Bob open a noiseless, authenticated, but insecure, public communication channel and communicate the photon arrival times and the basis in which each photon was detected. Of all coincidence events in which Alice and Bob simultaneously measured a photon they keep only those for which they both used the same polarization basis. After this basis reconciliation step, Alice and Bob both hold the sifted key. This bit string may still contain errors, due to experimental imperfections or due to eavesdropping, requiring a classical error correction procedure, followed by a process known as privacy amplification that further suppresses any information a hypothetical eavesdropper may have obtained. Only at the end of this step do Alice and Bob share a quantum secured secret key.

Here we report on a recently completed feasibility study towards the demonstration of optical quantum communication in free space between an Optical Ground Station (OGS) and a nanosatellite. By placing the entangled photon source on the ground the space segment contains the "Bob" detection system only, and therefore consumes less power, becomes smaller and less complex, thus increasing its reliability, and implementation in the 12U CubeSat standard is possible [10]. The space segment payload is also versatile: the receiver is compatible with multiple QKD protocols and other quantum physics experiments. In addition, the sensitive single photon detectors in combination with a small field-of-view telescope can be used to map light pollution on Earth at the quantum channel wavelength. This is important information for deciding the location of future optical ground stations that ideally would not be far from high population density, urban areas. The drawback is an increased, but still acceptable, effect of atmospheric turbulence on the link budget due to the shower curtain effect [11]. But this disadvantage of a higher uplink loss (by roughly 10 dB) is accompanied by the advantage of a lower photon detection rate on board of the satellite and therewith a significantly smaller amount of data to be stored and exchanged with the OGS via a classical (RF or optical), authenticated but non-secure communication channel.

In addition to its principal scientific aim of *demonstrating ground-to-space QKD with a CubeSat*, the NanoBob mission has the technological aims of:

a) Accurate clock synchronization between the ground-based station and the flight platform.
b) Fine attitude determination and control to ensure correct pointing of source and receiver under dynamic conditions.
c) The use of eye-safe laser beams at 1550 nm on the ground station and the space segment as laser tracking beacons, at the same time as they are used for fast classical optical communication using pulse position modulation, potentially at rates up to roughly 1 Gbit/s.

In the following we briefly discuss some relevant developments in the field before presenting a mission overview, the design of the NanoBob space segment, and the expected link budget. We limit ourselves here to the QKD mission scenario. Aspects relating to the duplex fast optical communication link, and the alternative mission scenarios of low light level dark area mapping and the quantum physics study of entanglement decoherence will be left for future reporting.

## 1.2. THE RACE TO SPACE: RELATION TO OTHER ONGOING PROJECTS

On August 16, 2016, the Chinese Space agency launched the 620-kg Micius satellite with on board the quantum communications experiment at space scale (QUESS) that includes an entangled photon-pair source. The payload is capable of establishing two simultaneous quantum downlinks to two ground stations on Earth 1200 km apart from a satellite that moves in a slightly elliptical orbit with an apogee at 584 km. The reported Bell test experiment showed that entanglement persisted over a combined distance of over 1600 km [12]. The same platform was used to demonstrate decoy-state QKD from satellite to a ground optical station near Beijing [13]. Also recently, researchers in Tokyo reported on a QKD experiment using a downlink from the 50-kg-class Socrates microsatellite [14], and the Singapore group operated an entangled source on a CubeSat [15]. Several other teams in Canada, Europe, and elsewhere, are working to bring quantum communication to space. Bedington et al. provide a table of notable satellite QKD proposals [16].

In 2002 first experiments using BB84 protocols were published demonstrating QKD on a horizontal link [17]. Experiments using entangled photons have been done over 144 km [18]. The losses experienced by the horizontal link through the turbulent atmosphere (~35 dB) are quite comparable to those expected for a single path between a ground station and a satellite in Low Earth Orbit (LEO).

To the best of our knowledge, NanoBob, having completed its end-of-phase-0 Mission Definition Review following ESA guidelines [19], is so far the most advanced European project focusing on the use of entangled photons and a CubeSat platform. It will demonstrate the feasibility of miniaturizing (both in volume and in power consumption) the Bob receiver module, promising to significantly lower the development time and cost of future quantum space missions, and opens the way to using a constellation of relatively cheap satellites to achieve global coverage and low latency.

NanoBob distinguishes itself by the use of a CubeSat receiver terminal that will be capable of executing most polarization-based single photon bi-partite protocols; most notably BB84 [5] and its more secure decoy-state variant [20], as well as the E91 protocol based on the Einstein-Podolsky-Rosen *gedanken experiment* [6]. Additionally other secure quantum communication tasks such as secure password authentication can be performed using the NanoBob payload and bit-commitment protocols [21]. Taking the expected link attenuation into account, we predict to be able to exchange keys of over $10^5$ bits during one OGS fly-over of the satellite (~3 min.). With such technology, one can already imagine an infrastructure arising, consisting of several optical ground stations that exchange quantum secure keys through a CubeSat in LEO (a trusted node) on a truly global scale. One satellite can consecutively exchange two different unconditionally secure keys with two different ground stations. A bit-wise XOR operation on the two keys on board of the satellite than yields a random bit sequence that can be shared publically with one of the two ground stations. This ground station can then compute the secure key held by the other ground station by repeating the same operation on the random bit sequence and its own secure key [22]. It is noted that whereas the Micius satellite node with its on-board entangled photon source does not need to be trusted as it exchanges a quantum key between two simultaneously visible OGSs, distributing a key on a global scale would require that the satellite reverts to the scheme mentioned above (and thus to become a trusted node), or otherwise, somehow, stores one of the entangled photons on board until it reaches the second OGS.

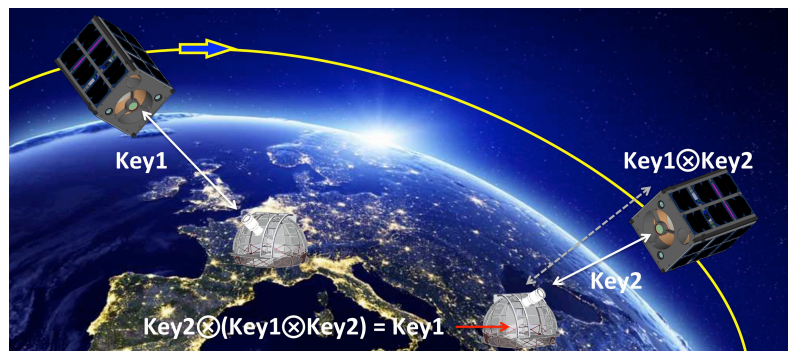

Figure 1. **Global unconditionally secure quantum key distribution** through a trusted node in uplink configuration [22].

In addition to demonstrating QKD in an uplink configuration, we prepare to use the beacon lasers required for the mutual tracking of the satellite and the OGS to establish an optical, two-way communication channel. Such a high-speed classical channel is practically mandatory for future satellite QKD operations if they are to exchange and negotiate useful (i.e., sufficiently long) sifted keys in a relatively short time frame. Using wavelengths in the telecomm region, as opposed to the visible or the short-wave infrared regions of the spectrum, implies that the classical communication channel becomes directly compatible with existing telecomm infrastructure.

## 1.3. QKD MISSION SCENARIO

The orbit in which the NanoBob satellite will be launched has to satisfy a number of criteria. First, it needs to comply with applicable space laws. The French law on space operations requires a decommissioning and destruction of the satellite upon its return in Earth's atmosphere within 25 years after end of operations [23]. For the 12U satellite without propulsion and with a weight of about 10 kg, this puts a higher limit of about 650 km to the height of a circular orbit. Several circular orbital scenarios were investigated using the Celestlab/Stela/VTS orbital simulation tools of the French National Space Agency (CNES). As the primary ground station location we use the ESA OGS at Tenerife on the Canary Islands (28.30086N, 16.51172W, 2410 m asl). We could satisfy the demand of maximizing the number of OGS encounters during nighttime by choosing an orbital inclination equal to the latitude of the OGS. However, as we want to be free to use other OGSs, and need to download data to RF ground stations located elsewhere, we conclude that a Sun Synchronous Orbit (SSO) at a height of 550 km and a local hour of 22h30 appears a near optimal choice. With an orbit time of 96 min, the satellite will make an average of 15 full orbits per day. Depending on the exact weight and the effective drag area (i.e., the product of the drag coefficient and the cross-sectional area perpendicular to the direction of motion), the expected lifetime is between 3 (10 kg, 0.11 m$^2$) and 11 years (24 kg, 0.08 m$^2$ – the surface area of one side panel). There are a fair number of rideshare launch opportunities into such an orbit, lowering the cost of the mission [24]. Limiting the distance at closest approach of the OGS at which the satellite passes to 750 km (i.e., the ground track passes within 500 km of the OGS), between 1 and 2 encounters per night can be expected, each with a total duration > 440 s (assuming tracking for elevations > 20°) of which roughly one half will be available for the QKD experiment. Figure 2 shows the ground tracks for the selected orbit.

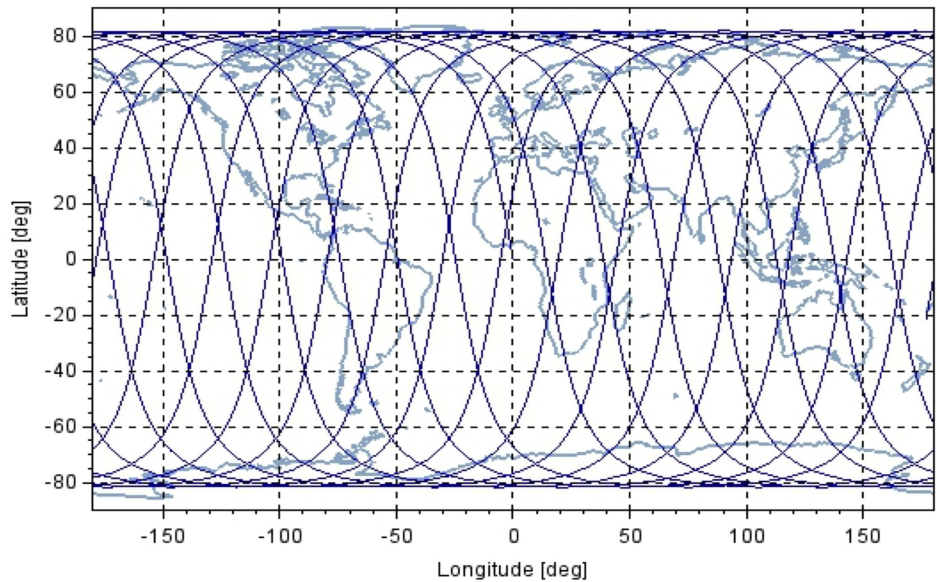

Figure 2. **Ground tracks** of the simulated SSO at 550 km altitude and with local hour of 22h30 over a period of 1 day. The inclination of the SSO orbit is 98°.

A typical encounter will have the satellite adapt its attitude just before arriving above the horizon, such that its telescope is oriented towards the expected location of the OGS. During this pre-acquisition flight segment pointing of the satellite towards the OGS relies on satellite ephemeris and star tracker data. A state-of-the-art integrated Attitude Determination and Control System (ADCS) designed for 6 to 12U CubeSats, such as, e.g., the XACT-50 by Blue Canyon Technologies, is already able to point the satellite with a 1-sigma precision of 50 µrad (11 arcsec) about an axis perpendicular to the star tracker bore (which in our case is parallel to the quantum channel line of sight). This should be sufficient to bring the OGS within sight of the satellite, given the FOV of 9 mrad (0.5°) of its beacon detection module, which images the ground laser beacon onto a quadrant detector, as well as onto a linear polarization analyzer. This will enable the satellite to fine-tune its attitude, both about the two axes perpendicular to the line of sight (using the quadrant signal) and about the line of sight (using the linear polarization of the beacon laser). Inertial calculations show that this process should not take longer than about 30 s. At the same time, the OGS beacon laser illuminates the corner cubes on the satellite and the satellite may turn on its beacon laser to make it more easily visible to the OGS. Either way the beacon light received by the OGS telescope will enable it to acquire, and start tracking, the satellite. The corner cubes are built-in as a back-up solution for the satellite beacon laser. They have the added advantage that they return the beacon laser towards the OGS, also if the satellite's telescope is not (yet) accurately oriented towards the OGS (using, e.g., star tracker information). This is in contrast to the satellite beacon laser, which will not be seen by the OGS until the satellite is fairly accurately directed towards the OGS. We note that the OGS will be equipped with a changeable dichroic beamsplitter in order to adapt the detection wavelength to that of the beacon laser being used (i.e., OGS versus satellite). Accurate pointing, acquisition, and tracking (PAT) during the quantum science segment of the flight over the OGS thus rely on direct feedback of error signals obtained from detection of the beacon lasers.

At this point, with satellite and OGS telescope tracking each other, the exchange of a quantum key can commence. During the next roughly three minutes the satellite detects and times the arrival of single photons that are collected by its telescope and analyzes their polarization state.

At the end of, or already during, this phase the satellite opens an authenticated public communication channel (either optical or RF, with the same or another ground station) and sends the photon arrival times and the basis in which each photon was detected to the ground station. The latter then proceeds with clock synchronization by performing a cross-correlation operation on the time series of photon detection times, comparing it to its own time series of photon detection events [18, 25]. This procedure reduces the coincidence time window to roughly a few hundred picoseconds, ultimately limited by detector jitter. A small coincidence time window reduces accidental coincidences due to detector dark counts and residual background counts. Finally, the ground station and satellite carry out the basis reconciliation, error correction, and privacy amplification steps to produce the quantum secure key shared by the OGS and the satellite. In section 1.7 we will provide an estimate of the rate at which such a key can be constructed.

The quantum channel will operate at a wavelength of 808 nm, a choice that reflects the availability of a highly efficient entangled photon source and of single photon detectors that combine sub-nanosecond jitter with a high quantum efficiency and that require only modest cooling in order to achieve a low dark count rate. At the same time, while atmospheric absorption and scattering are higher at this wavelength than in the telecom wavelength range, these effects are more than compensated for by the relatively high photon detection efficiency.

During the daylight part of the orbit the satellite orients the solar panels on one or two of its sides towards the Sun in order to recharge the batteries. The use of deployable solar panels is avoided as their limited rigidity could reduce the precision of the satellite's pointing. The star-tracker and telescope sun exclusion angles (~45 degrees) are automatically satisfied in this orbital scenario, while it is also compatible with communication with an RF ground station, as the S-band patch antennas will be located on the opposing side panels.

## 1.4. CRITICAL SATELLITE SUBSYSTEMS

The NanoBob mission will miniaturize the Bob receiver payload for it to fit inside a 12U CubeSat frame. This size limit is chosen as the smallest CubeSat standard that allows for a reasonably large main telescope of 150-mm diameter (potentially up to 180-mm diameter), increasing the light collection efficiency by a factor of four (6 dB) compared the alternatives of 3U or 6U and providing sufficient space to incorporate the required beacon laser with a secondary, smaller telescope. Figure 3 gives a schematic representation of the assembly, while Table 1 gives the estimated size, weight, and power (SWaP) consumption of the subsystems together with their uncertainties. The definition of the Technology Readiness Levels is according to ISO standard 16290:2013 as adopted by ESA [26].

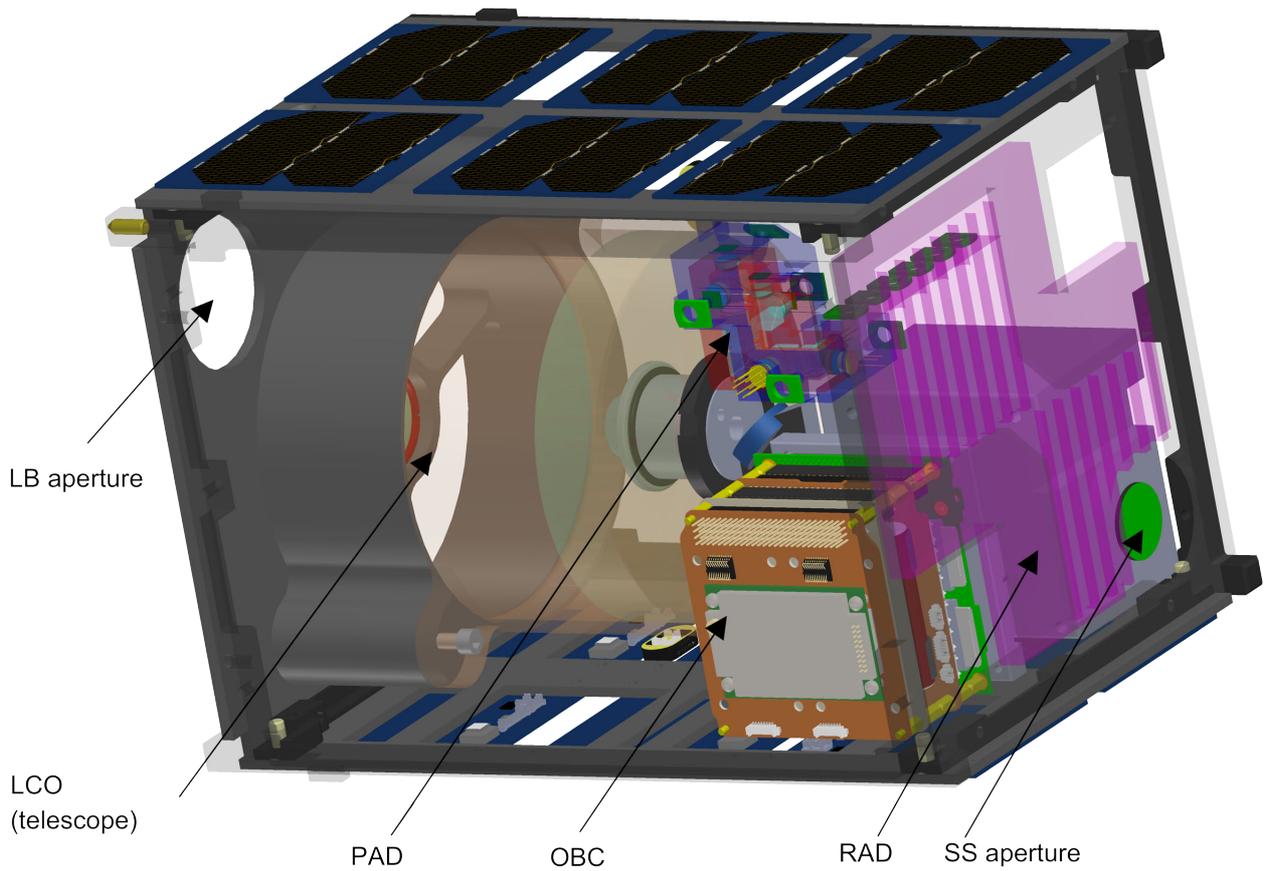

Figure 3. **Computer aided design of the assembly** showing the main components of the light collection optics (LCO), polarization analysis and detection module (PAD), star tracker (SS), on-board computer (OBC), solar panels and the radiator (RAD) for passive detector cooling. Notably, the laser beacon (LB) with its telescope and the battery packs are not shown to provide an unobstructed view of the remainder.

Table 1. Results of a SWaP (Size, Weight, and Power) analysis, including contingency.

| Item | Size (ml) | Weight (g) | Peak Power (mW) | TRL | Margin | Size Margin | Weight Margin | Power Margin |
|---|---|---|---|---|---|---|---|---|
| **Payload** | **5 045** | **2 680** | **14 500** | | | **1160** | **819** | **6850** |
| Quantum Optical Module (808 nm) | 125 | 200 | 4 000 | 4 | 50% | 63 | 100 | 2000 |
| Beacon Receiver Module (1530 nm) | 145 | 360 | 6 000 | 3 | 50% | 73 | 180 | 3000 |
| LCO-QKD | 4 050 | 830 | 0 | 4 | 20% | 810 | 166 | 0 |
| Beacon Transmitter Module (1565 nm) | 200 | 350 | 1 500 | 2 | 50% | 100 | 175 | 750 |
| Retro-reflector | 125 | 340 | 0 | 7 | 20% | 25 | 68 | 0 |
| Detector cooling | 100 | 300 | 0 | 2 | 20% | 20 | 60 | 0 |
| Time Tagging Module | 100 | 100 | 2 000 | 2 | 50% | 50 | 50 | 1000 |
| Beacon Signal Processing | 100 | 100 | 500 | 7 | 10% | 10 | 10 | 50 |
| Data storage | 100 | 100 | 500 | 7 | 10% | 10 | 10 | 50 |
| **Platform** | **5 425** | **5 148** | **12 060** | | | **403** | **443** | **617** |
| OBC | 110 | 94 | 500 | 9 | 5% | 6 | 5 | 25 |
| ADCS | 750 | 1 225 | 2 470 | 9 | 5% | 38 | 61 | 124 |
| GPS | 35 | 24 | 1 200 | 9 | 5% | 2 | 1 | 60 |
| UHF/VHF module | 110 | 75 | 4 000 | 9 | 5% | 6 | 4 | 200 |
| S-Band module | 130 | 62 | 3 800 | 9 | 5% | 7 | 3 | 190 |
| Antennes | 110 | 128 | 0 | 9 | 5% | 6 | 6 | 0 |
| PMU & batteries | 680 | 840 | 90 | 9 | 20% | 136 | 168 | 18 |
| Mechanical structure | 3 000 | 2 000 | 0 | 9 | 5% | 150 | 100 | 0 |
| Detector radiators | 200 | 400 | 0 | 5 | 20% | 40 | 80 | 0 |
| Solar panels | 300 | 300 | 0 | 9 | 5% | 15 | 15 | 0 |
| **TOTAL PAYLOAD & PLATFORM** | **10 470** | **7 828** | **26 560** | | | **1563** | **1262** | **7467** |

The SWaP analysis of Table 1 shows that the estimated maximum volume including contingency is 12 L, the maximum weight is 9 kg, and the peak power consumption can reach 34 W. Both volume and weight are well within the limits of 19.9 L and 24 kg imposed by the 12U CubeSat standard [10]. Table 1 also enables estimation of the energy consumption per orbit. The most critical orbital scenario is, not surprisingly, the scientific scenario of a QKD experiment. For a worst case estimation we assume that the initial alignment phase takes 5 minutes, the quantum experiment lasts 5 min., the beacon lasers will be operated during this entire period (10 min.) and the S-band communication with the ground station lasts 10 min. We then calculate an energy consumption of 9.2 Wh during one full orbit. This is to be compared with the recharging capacity of the batteries of 21.6 Wh provided by the solar panels during the same orbit. It also means that the installed battery capacity of 66 Wh will see a cycling of less than 15% of its nominal, initial capacity. We thus expect that the batteries can easily sustain the ~16400 cycles during the longest expected operational lifetime of the satellite of 3 years (which is more likely limited by radiation damage to the detectors).

In the following sections we describe the subsystems that have been identified as most critical to the mission outlined above. All other subsystems (such as power systems and OBC) can be purchased commercially off-the-shelf and be used with minimal modification. They also generally have space heritage.

### 1.4.1. LIGHT COLLECTION OPTICS

The optical module (see Fig. 4) is literally (at) the center of the payload. It consists of a telescope with high light gathering power followed by the quantum channel polarization analyzer and a separate unit dedicated to detecting the ground-to-satellite beacon laser. It is complimented with a small diameter telescope that focuses the satellite beacon laser, as well as two corner cubes that retro-reflect the OGS beacon laser.

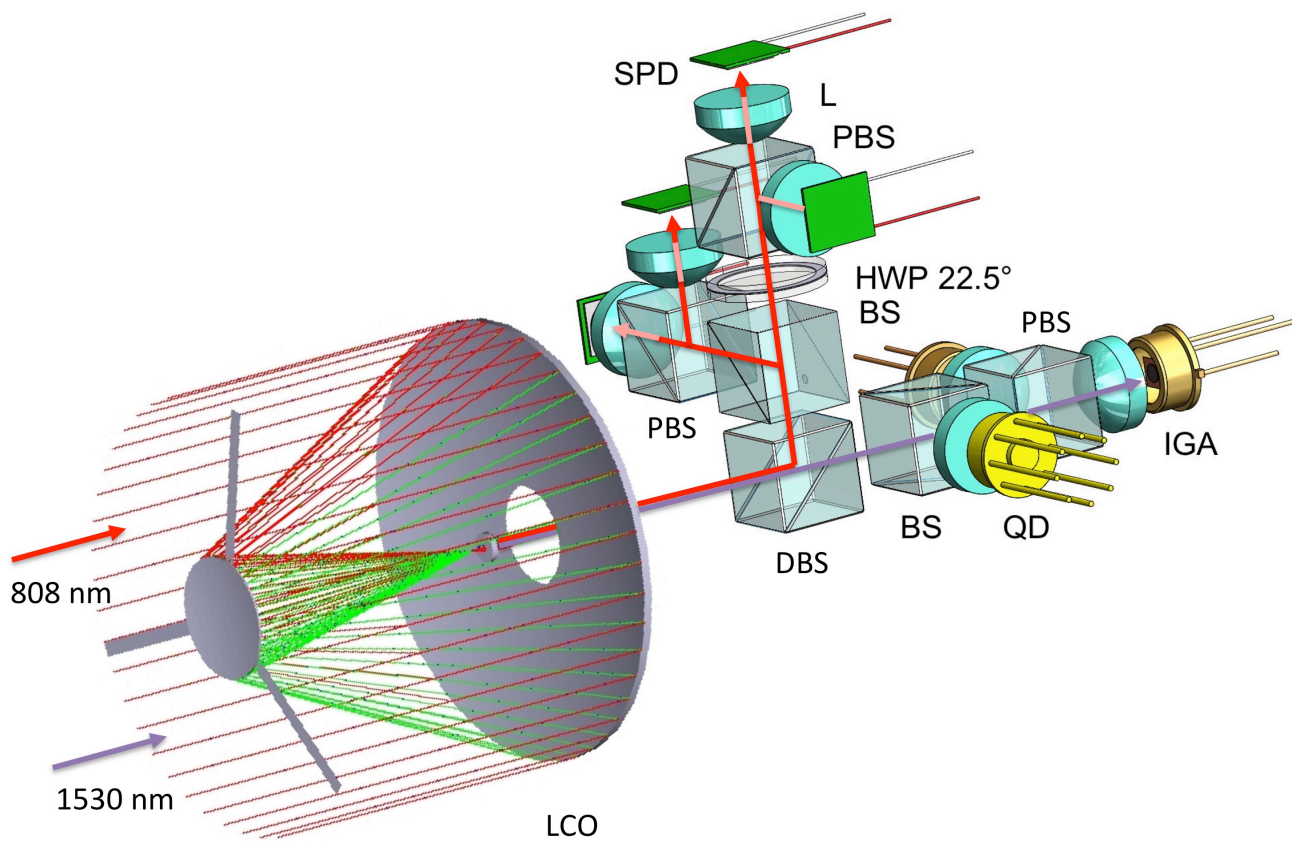

Figure 4. **Schematic representation of the optical module**. The OGS beacon laser at 1530 nm is collected by the main telescope (LCO). After separation by a dichroic mirror (DBS) it is split, with part being send to the beacon polarization analyzer (consisting of a polarizing beam splitter (PBS) and two detectors), and part being focused onto a quadrant detector (QD). The quantum channel light at 808 nm collected by the main telescope is sent towards a 4-detector polarization analyzer that includes two polarizing beam splitters (PBS), one for the {HV} basis, the other for the {DA} basis, and one half-wave plate (HWP) that rotates the polarization by 45° for the {DA} basis. The {HV} versus {DA} basis choice occurs randomly in the beam splitter (BS). Not shown are the corner cubes that retro-reflect the OGS beacon laser at 1530 nm, and the small diameter telescope that directs the satellite's beacon laser at 1565 nm towards the OGS. The telescope and the polarization analyzer/detection module are not at the same scale.

The light collection optics should maximize the number of photons captured from the photon stream directed towards the satellite by the OGS. Ideally, the OGS produces a diffraction limited beam diameter of a little over 1 meter at the location of the satellite for the 808 nm quantum channel; in practice increased to several meters due to atmospheric turbulence. Increasing the receptor aperture will directly result in higher signal. Losses internal to the quantum channel light collection optics and the polarization analysis module should also be minimized. The receiver telescope must preserve the polarization direction of the incoming photons, such that it contributes not more than 0.25% to the total polarization error (see section 1.4.3). This signifies that the receiver telescope is polarization neutral to the extent that the spread in polarization of beams taking different paths through the telescope will be less than 1°. Starting point for the optical design is a Cassegrain telescope with an opening aperture of 150 mm diameter and an overall length of just 125 mm. A refractive solution was not considered. Although the weight of a lens system could be reduced using a Fresnel lens, strong accelerations along the optical axis expected during launch are a serious concern, as is radiation damage of the optics.

The Field Of View (FOV) of the quantum channel's detectors (100-µm diameter) should in practice be as small as possible while respecting the constraint of the dynamic pointing stability of the pointing and tracking system (see section 1.4.4). This in order to reduce unwanted background light from being captured by the receiver telescope. Considering this, the quantum channel FOV is 215 µrad (45 arcsec), corresponding to a circular footprint of 120 m diameter with the satellite at an orbital height of 550 km. Knowledge of the photon intensity or spectral radiance of the area of the OGS then enables one to calculate the expected background count rate. The Vienna group made measurements at the Canary Islands with a spectral band pass filter of 10 nm centered at 810 nm, resulting in a photon flux of $10^{10}$ to $2.5 \cdot 10^{11}$ $s^{-1}sr^{-1}m^{-2}$ depending on the moon phase [27]. Even at a distance of 1100 km between OGS and satellite, near the beginning and end of their encounter, the background count rate is then still smaller than 400 cps (counts per second), given a 15-cm receiver telescope diameter and taking an atmospheric attenuation of ~3 dB into account; acceptable for a Bell test with uplink losses < 50 dB (cf. the calculated Visibility of Figure 9). We note that the actual background can be further reduced using bandpass filters with a narrower transmission profile; 3 nm appearing a reasonable choice for which center wavelength transmission of >90% is still possible and outside bandpass blocking is better than OD6 (60 dB).

In order to compact the whole instrument while conserving a small ratio of the diameters of the secondary and primary mirrors (i.e., a better transmission), a relatively high field curvature has been chosen. Considering the on-axis aberrations, we take benefit of the Cassegrain design, which enables totally suppressing the spherical aberration (SA3) by choosing the conic constant (also known as the Schwarzschild constant) of the hyperbolic secondary mirror. Aberrations are in general not critical given the small FOV and the non-imaging character of the application. In particular, the aberrations appearing within the FOV (coma, FOV curvature, distortion) can be neglected. The design was analyzed in ray tracing software to show that a 100-µm diameter photodetector behind the telescope can capture more than 80% of the incoming light intensity.

The FOV of the beacon detection is 9 mrad (see section 1.3). The compact telescope allows for the entire optics module to be shorter than 200 mm.

### 1.4.2. POLARIZATION ANALYSIS

The polarization detection unit analyzes the incoming photons in either one of two bases (see Figure 3). An easy and secure way to make the random choice of selecting either one of the two bases is by the use of a 50/50 beam splitter (BS) [28]. As pointed out by Gisin et al. [7], the quantum mechanical nature of the underlying physical process guarantees its randomness, but experimental artifacts, notably detector dead-time, afterpulsing, and detector flashes [29] could potentially lead to correlated adjacent bits at high photon rates [30, 31, 32]. Following the BS a half-wave plate (HWP) oriented at 22.5° in one of the two paths is used to rotate the polarization direction by 45°. Polarizing beam splitters (PBS) in both paths enable the polarization analysis. The polarizer extinction ratio and the orientation/mounting precision of the PBS are such that the probability of a photon ending up in the wrong path (e.g., a vertically polarized photon being detected by the "horizontal detector" instead of the "vertical detector") is not larger than 1%, as such a detection error ($e_d$) increases the coincidence error and therewith reduces the signal-to-noise ratio and visibility (section 1.7). Importantly, this error includes the possible misalignment of the OGS and satellite polarization bases.

All quantum communication protocols based on polarization encoding of the qubits require a shared reference frame between the transmitter (Alice) and receiver (Bob). Atmospheric turbulence, scattering, and the Faraday effect can potentially rotate the plane of polarization. It is, however, easily shown that these effects are negligible (< 1 mrad) compared to geometrical effects due to the moving satellite and the moving mirrors of the transmitter telescope. The latter effect was studied by Bonato et al. [33] and should be compensated by appropriate rotation of the polarization bases of the OGS or satellite. If these bases would be misaligned by 4°, this would contribute 0.48% to the detection error. Two options are available: The first is to rotate the OGS polarization basis (e.g., by the motorized rotation of a half-wave plate (HWP) in the quantum light channel) to adapt to the satellite orientation. The latter is known to the OGS from the pre-programmed flight plan and the information received at regular intervals (~100 ms) from the satellite's star tracker measurements. Fine-tuning will take place using a signal obtained from the analysis of the linear polarization of the satellite's beacon laser as received by the OGS [34]. A second option entails rotation of the satellite about its seeing axis using an error signal derived from analysis of the separately controlled linear polarization of the OGS beacon laser, again combined with data from the star tracker. Both solutions avoid addition of moving parts (the rotatable HWP) to the satellite. We fully implement the first solution, but equip the satellite with the hardware required for the second option. In case of failure of the first option, for example due to a satellite beacon laser failure, the satellite can be re-programmed to implement the second solution. Even though the dynamic tracking precision of the ADCS is generally significantly worse about its star tracker bore axis (which is parallel to the receiver telescope seeing axis), it is however more than sufficient to allow precise pre-orientation of the satellite about its seeing axis (see section 1.4.5). The OGS laser beacon signal is then used to improve absolute accuracy and to further improve alignment precision to the 10-µrad level. Ground-based experiments will verify that the OGS laser beacon polarization correctly tracks the orientation of the OGS polarization bases.

The coincidence count rate shows a $\cos^2$-dependence when varying the measurement basis between HV and DA. The visibility of this polarization correlation decreases, not only due to the above mentioned polarization detection error, but also due to source imperfections, polarization imbalance in the quantum link, and detector dark and background counts (see section 1.7).

### 1.4.3. SINGLE PHOTON DETECTORS

Based on the link-budget and key rate analyses presented in sections 1.6 and 1.7 we require the single photon detectors (SPDs) to have a photon detection efficiency (PDE) > 40%, dark count rate (DCR) per detector < 1000 cps, timing jitter < 100 ps, afterpulsing < 3%, and a maximum count rate >100 kHz without saturation effects. Afterpulsing will contribute to the dark- or background count rate, and may also lead to a correlation between bits. The light collection optics have been designed for a detector diameter of 100 μm.

The wavelength of operation is not a primary specification. Two wavelength ranges appear potentially attractive for free-space QKD: the near-infrared region near 800 nm, and the telecom, short wave infrared (SWIR) range around 1550 nm. The link budget slightly favors the longer wavelength (see section 1.6). Since key distribution and the sending of encrypted messages are in principle independent aspects of cryptography, there is no fundamental reason to operate the QKD channel on the same wavelength as that used for a fiber-based network used to transmit the encrypted message. That said, if operating at telecom wavelengths the quantum signal could be directly transmitted by fiber from OGS to "client" for polarization analysis and detection. There also remains an obvious interest in mutualizing optical building blocks between the free-space and fiber-based systems, which drives the exploration of the feasibility of QKD at 1550 nm. However, currently, neither of the available detector technologies in the 1550 nm region is attractive for use in a CubeSat: Both Indium-Gallium-Arsenide (IGA) APDs as well as detectors based on Mercury-Cadmium-Telluride (MCT) technology require cooling to very low temperatures (< -80 °C). In addition, IGA APDs have a rather low photon detection efficiency (PDE) < 25%, whereas MCT SPDs are still in development and appear to be hampered by large DCR [35, 36]. At the current state of technology, only Silicon-based Avalanche Photo Diodes (Si-APDs) in the 800 nm range are able to combine a sufficiently high PDE and low jitter with a low DCR. Si-APDs have been operated and characterized in space or under space radiation conditions. This has clearly shown the need for special measures to keep the dark count rate below acceptable levels, also after longer times in a space environment [37, 38, 39]. To our knowledge, no similar space heritage exists for IGA SPDs, let alone MCT SPDs.

The Si-APD that was identified for use in the NanoBob quantum channel is manufactured by Micro-Photon Devices. In particular, the Red-Enhanced version of this detector shows an improved sensitivity towards 800 nm (PDE = 40%) and is also very attractive as it combines a low reverse voltage (50 V) with low jitter (90 ps) and dark count rate [40]. Additionally, the specified low dark count rate of 25 cps was demonstrated at a temperature of -5 °C, much higher than the -30 °C targeted in our system. We expect to receive prototypes of these detectors shortly for radiation testing in Grenoble.

We note that the DCR requirement has obvious implications for the detector operating temperature. However, the stability of the detector temperature is not very critical for the QKD experiment, but may limit the precision and accuracy that can be attained if the space segment is to be used in light pollution mapping mode (see section 1.1). High doses of radiation in space may cause the DCR to increase over time. For this reason the detectors are shielded by housing them in an aluminum module with walls of minimally 10-mm thickness, as well as by other satellite components around it (batteries, electronics, the aluminum CubeSat structure, and solar panels). Using the OMERE software package [41] we calculated the cumulative total radiation dose received by the detectors as a function of the thickness of the aluminum shielding provided by the mounting structure. The satellite was assumed to be in an SSO at 550 km with a launch date in September 2020. The results for a 1-year and a 3-year exposure are shown in Figure 5. The total incident radiation dose includes contributions from electrons trapped in Earth's magnetic field, solar and trapped protons, and Gamma photons (in order of decreasing radiation level). Kodet et al. [42] determined that gamma radiation has no detrimental effect on Si-APD performance, and in any case in our mission scenario the gamma radiation dose accounts for just 1‰ of the total. Anisimova and colleagues tested several different Si-APDs shielded by 10 mm of aluminum under similar radiation exposures and found the DCR of the small area to increase to several hundred cps [39]. Packing the detector unit in a hydrogen-rich material such a polyethylene may further reduce the total radiation dose. This will be part of the radiation testing of the above-mentioned prototype detectors.

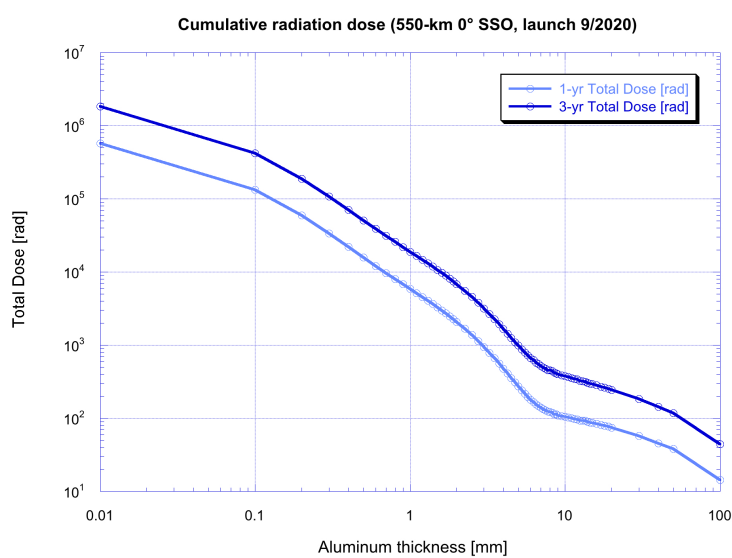

Figure 5. **The effect of radiation shielding** by the aluminum detector housing on the total cumulative radiation dose for exposures of 1 and 3 years as calculated by the Omere software package [41]. The satellite is assumed to be in a 550-km SSO starting September 2020.

It has been shown that annealing of Si-APD detectors at elevated temperature (60 to 100 °C) for several tens of minutes can already lower the dark count rate significantly (up to about an order of magnitude decrease) [37, 39, 42]. For this reason, it may actually be advantageous to let the detectors heat up during the daytime part of the orbit.

In fact, the detectors will be cooled passively during the nighttime part of the orbit using a radiator facing deep space. Small local heaters will be used to regulate the individual detector temperatures to -30+/-1 °C. We thus do not use thermo electric cooling (TEC) of the detectors, also not for final stage cooling or as a temperature fine-tuning solution. This comes with some notable advantages: TEC units are notoriously inefficient with a low coefficient of performance. More problematic appears the risk of total or partial failure of the TEC or its power supply, in which case the TEC unit would act as a thermal insulator between the detector chip and the mounting structure. The TEC unit would also introduce a mechanically less rigid element that may affect detector positioning. Relying solely on passive cooling and low-power resistive heating thus increases the reliability of the detector thermal management system.

To study the passive cooling of the detector module in some detail we modeled the spacecraft as a square cuboid of size 22 x 22 x 34 cm$^3$. Its panels are covered with a multi-layer insulation (MLI) characterized by an IR emissivity of 0.71 and a UV absorptivity of 0.52, whereas the radiator is coated white with an IR emissivity of 0.81 and UV absorptivity of 0.25. The average spacecraft temperature in a 550 km SSO is taken to be 10 °C. The detector unit is modeled as an aluminum block connected to the radiator with a thermally conductive strand with a total resistance of 3.2 K/W. Each of the four detectors and its proximity electronic circuitry consumes 0.3W. The incoming direct solar UV/VIS radiation, the reflected radiation from Earth's surface, and Earth's emitted IR radiation during a typical QKD orbital scenario with nighttime OGS encounter was calculated using Airbus' Thermica software [43]. Taking further into account the different radiative and conductive heat fluxes between the satellite structure, the radiator, and the detector unit, the model developed allows us to calculate the minimum radiator surface area needed to maintain the detector module temperature below -30 °C. Depending on whether the radiator is placed on the square end-panel facing deep space (the panel that also accommodates the star tracker) or on one of the space facing side panels, the calculated required surface area varies between 0.052 and 0.055 m$^2$. In practice the radiator area will be distributed over the end-face and one or two side panels. Maximizing the radiator area to the available 0.19 m$^2$ may enable cooling of the detectors to a lower temperature still. This is clearly favorable in light of the recent findings that show that deep cooling drastically reduces and even mitigates the effects of radiation [39].

### 1.4.4. TIME TAGGING

The events detected by the Bob quantum receiver can be due to detector dark counts, background (stray) light, or the entangled photons sent by the OGS. Identification of the entangled photons is done by comparing their time of arrival at the NanoBob quantum receiver with the arrival times of the other photon of the entangled pair at the Alice detection unit at the OGS. Such identification through coincidence timing requires a high timing precision if large numbers of photons are involved. With a source single photon generation rate of 100 Mcps, a timing resolution (coincidence time window) better than about 1 ns is required in order to reduce the probability of accidental coincidence to an acceptable minimum. A better timing resolution will thus increase the signal-to-noise ratio (see section 1.4.2) by suppressing the number of background or dark counts being accidentally registered as an entangled photon event.

In order to time stamp the photon arrival a time-tagging module is used, both at the OGS [44] and on the CubeSat. An integrated space-qualified system will be specifically designed using a dedicated integrated circuit implementing time-to-digital conversion (TDC). A short-term stability of the TDC oscillator of 0.1 ppb (10$^{-10}$) is required, corresponding to a measurement precision of about 10 ps for an average time between photon arrivals that could be as long as roughly 100 ms (10 cps). This can be achieved using an oven controlled crystal oscillator (see, e.g., [45]) or miniature atomic clock (such as, e.g., the model Quantum SA45.s by MicroSemi [46]). Long-term clock synchronization between OGS and satellite is then achieved by the fore-mentioned time correlation technique applied repeatedly on data over intervals of approximately 100 ms [25, 44]. Implementing TDC with a time resolution < 25 ps and jitter < 10 ps in integrated circuitry is challenging but can be done in standard field programmable gated arrays (FPGAs) using a method based on self-timed rings (STR) [47]. Alternatively, Vernier-TDC will be employed if the compact STR-based approach turns out to be too difficult to implement in an FPGA.

The combined contribution to the coincidence time window of the detector and electronics jitter on the space segment, and those of a state-of-the-art OGS [44], is about 100 ps.

### 1.4.5. POSITION, ACQUISITION AND TRACKING

A first concern for the PAT of the satellite is whether the precision of its ADCS is sufficient, also under dynamical conditions. For a circular orbit at an altitude of 550 km the slewing rate required to keep the line of sight of the satellite along the line segment from OGS to satellite is reaches a maximum value of ~12.6 mrad/s = 0.72°/s at closet approach (0° zenith angle). The slewing rate required of the OGS telescope to track the satellite reaches a maximum value of 13.7 mrad/s = 0.79°/s. These values are compatible with OGS telescopes designed to track LEO satellites, such as the ESA OGS "Observatorio del Teide" at Tenerife, situated at an altitude of 2.393 m, and also less stringent than the capabilities of the best commercial CubeSat ADCSs.

The current demonstrated state-of-the-art in terms of attitude determination and control appears to be held by the XACT family of ADCS manufactured by Blue Canyon Technologies [48]. Their XACT-15 module was integrated in the MinXSS 3U CubeSat [49], launched December 6, 2015 and the RAVAN 3U CubeSat [50], launched November 11, 2016. On MinXSS it has demonstrated to exceed its specifications of a pointing accuracy < 50 μrad (11 arcsec) and a pointing knowledge < 30 μrad (6 arcsec) (both 1-sigma) for the two cross-star tracker-bore sight axes. The pointing accuracy about the bore axis is specified to be < 120 μrad (25 arcsec). Furthermore, the dynamic tracking error (1-sigma) of the XACT unit as a function of the slewing rate for the two cross axes is largely unaffected for slewing rates < 1.1°/s. Even the dynamic tracking error about the bore sight axis does not exceed 480 μrad (100 arcsec), which is still well within our requirements.

For the Blue Canyon XACT-50, which is identical to the XACT-15, except for its larger 50 mNms reaction wheels, to guarantee a slewing rate of at least 1 °/s in any axis, the moment of inertia in the slewing axes needs to be below 2.8 kgm². The predicted moments of inertia of the NanoBob satellite are about one-twentieth of this value.

At a satellite altitude of 550 km it takes the beacon laser photons at least 1.83 ms to arrive at the satellite. During this time the angular displacement of the satellite, as seen from the OGS telescope position, could be as much as 25 μrad. This is non-negligible with respect to the telescope FOV and will have to be taken into account in its tracking control. Similarly the satellite attitude will need to slightly point ahead of the acquired OGS position.

### 1.4.6. BEACON LASERS

Knowledge of the attitude (orientation) of the satellite is typically limited to about 50 μrad by star tracker performance. While this is almost an order of magnitude smaller than the satellite's quantum channel FOV, this may not be sufficient for accurate pointing due to ephemeris uncertainty that limits the ability to accurately transfer the attitude knowledge in the inertial frame to the Earth-fixed frame. On the other hand, the OGS requires accurate knowledge of the satellite position in the Earth-fixed frame in order to accurately track the satellite. For the same reason as before, data from the star tracker may not be precise enough. The positioning error of a Commercial-Of-The-Shelf (COTS) GPS receiver can be as large as 10 m [51], even though sub-meter precision has been shown on a LEO spacecraft [52]. This, however, could already put the satellite out of sight of the OGS, considering that even in the presence of atmospheric turbulence a 1-m diameter telescope would illuminate a disk with a diameter of just a few meters at the altitude of the satellite.

To provide an additional, and more accurate way to align both the OGS telescope and satellite receiver we will implement a two-way beacon (guide star) system, allowing for relatively fast closed-loop control of the satellite attitude, as well as satellite tracking by the OGS telescope. The beacon receiver module on the space segment includes a quadrant detector (or alternatively or CCD camera) to enable attitude control about the two axes perpendicular to the line of sight, and a linear polarization analyzer made up of a polarizing beamsplitter and two IGA photodetectors.

The initial choice of wavelength for the beacon lasers is in the NIR C-band around 1550 nm as here efficient lasers and detectors are easily available and the atmospheric transmission is high. Moreover, the wavelength is retina-safe, and directly compatible with existing telecommunication hardware and infrastructure. It is also advantageous that optical communication in space has been demonstrated previously in this wavelength range [53]. We therefore aim to use the beacon lasers not only for PAT, but also for fast optical communication by implementing a pulse position modulation scheme [54]. Optical communication provides an attractive alternative to RF communication by virtue of its lower power demand and high data rate.

The use of a beacon laser and optical communication using a laser beam between a ground station and a LEO CubeSat have been separately investigated by other groups [55, 56, 57]. We will implement a very similar design as those explored by the groups mentioned here, and DLR in particular [55]. It should be noted that the uplink experiences a higher link loss (by about 10 dB) due to atmospheric turbulence, but that this could be compensated by the use of a higher power laser. The downlink experiences lower losses and the OGS can be equipped with a large diameter receiver (or use the Coudé focus of the main telescope) as well as cooled high-sensitivity detectors, together allowing for the use of a relatively low power laser source and small transmitter telescope on the space segment. Finally, we note the encouraging result reported in [58] that the large difference in quantum channel and beacon laser wavelength does not preclude using the beacon laser at 1550 nm to correct the turbulence-induced beam wander at the quantum channel wavelength of 808 nm (employing, e.g., a fast steering mirror on the OGS or its adaptive optics system [59]).

## 1.5. GROUND STATION AND ENTANGLED PHOTON SOURCE

A number of telescopes that satisfy the needs of the experiment have been identified [27]. The most promising of these is the ESA OGS at Tenerife. Equipped with a 100-cm telescope, it is capable of tracking a satellite in LEO with a pointing precision of 1.2 μrad starting at relatively low elevation angle (~15°). In order to characterize the strength of the atmospheric turbulence above the telescope, the Fried parameter $r_0$ (a.k.a. the atmospheric coherence width) [60] has been measured at the RoboDIMM ORM telescope on the Canary Islands over a 8.5-year period, showing that, on average, about 112 days per year $r_0 > 20$ cm ($\lambda$=810 nm) (Fig. 6) [44]. The OGS telescope aperture is in fact generally larger than the average Fried parameter for the location of the OGS, such that the beam size at the position of the satellite is not diffraction limited, but rather limited by atmospheric turbulence.

The optical ground station will be equipped with an entangled photon source and the associated Alice detection module to enable the implementation of the E91 QKD protocol with the qubits encoded in linear polarization states of the photons [44]. The experiment is based on photon pairs produced by spontaneous parametric down conversion (SPDC). This nonlinear process consists of splitting one photon with energy $h\nu_p$ into two lower energy photons at $h\nu_s$ (signal) and $h\nu_i$ (idler) inside a nonlinear crystal exhibiting a strong second-order electric susceptibility $\chi_{(2)}$. The pair of photons that is created can exhibit entanglement when they are indistinguishable in terms of their momentum vectors. SPDC is not very efficient. The Vienna source can generate up to about $8\cdot 10^6$ pairs per second per mW of pump power, for a maximum pair generation rate of $3\cdot 10^8$ s$^{-1}$ [4]. Improving the brightness of the source would enable increasing the key rate of the QKD protocol (see section 1.7).

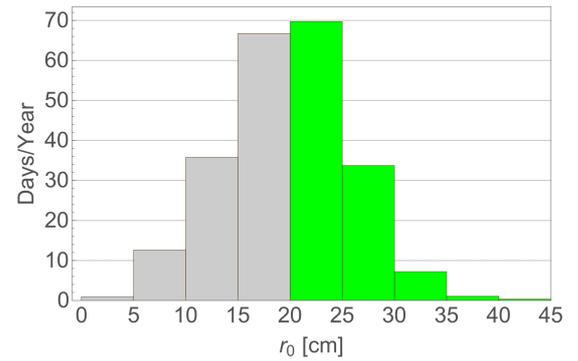

Figure 6. **Histogram of the Fried parameter at 810 nm**, based on observations from Jan. 2, 2009 to April 22, 2017 at the Observatorio del Roque de los Muchachos (ORM) at La Palma [44]. The histogram includes 228 days per year; during the remaining 137 days no measurements were possible due to overcast or technical problems. During 112 days/year $r_0 > 20$ cm (green area in the histogram).

## 1.6. LINK BUDGET

We estimate the average link attenuation between the OGS and the satellite receiver using the following formula [61]:

$$A = \frac{L^2(\theta^2_T + \theta^2_{atm})}{D^2_R} \frac{1}{T_T(1-L_P)T_R} 10^{\frac{A_{atm}}{10}} \quad (1)$$

Here, $L$ is the link distance between the OGS and the satellite, $D_R$ is the receiver diameter, $T_R$ and $T_T$ are the transmission factors of the receiver and transmitter telescopes, respectively. $L_P$ is the pointing loss due to misalignment, and $A_{atm}$ is the atmospheric attenuation due to (Rayleigh) scattering and absorption (expressed in dB) that is a function of the path length through the atmosphere and thus the zenith angle $\zeta$: $A_{atm} = A_{atm,0}(L/h) \approx A_{atm,0}/\cos(\zeta)$, where $h$ is the height of the satellite orbit, $A_{atm,0}$ equals 3 dB at 808 nm and 2 dB at 1550 nm. The angles $\theta_T$ and $\theta_{atm}$ are respectively the diffraction limited and atmospheric turbulence induced divergence angles of the transmitter telescope that are assumed to add quadratically. We define these two "seeing" angles as follows:

$$\theta_T = 2.44 \frac{\lambda}{D_T} \quad (2)$$

and

$$\theta_{atm} = 2.1 \frac{\lambda}{r_0} \quad (3)$$

The definition of $\theta_T$ differs from the one given Pfennigbauer et al. [61], who used $\theta_T = 1.22\ \lambda/D_T$. Since we do not want to underestimate the effect of atmospheric turbulence, we use the definition of Eq (2), such that $L\cdot\theta_T$ corresponds to the full diameter of the central spot in the Airy diffraction pattern (defined by the first zero-crossing of the Airy function), instead of its radius.

For the same reason we use the original definition of Eq. (3) for $\theta_{atm}$, even though some authors (including [60]) have used $\theta_{atm} = \lambda/r_0$, thus without the factor of 2.1, which equals the ratio of the spatial coherence radius $\rho_0$ to the Fried parameter $r_0$ [60].

The Fried parameter $r_0$ corresponds to the diameter of the diffraction limited telescope in the absence of atmospheric turbulence that would yield the same resolution as a telescope with a diameter much larger than $r_0$ but in the presence of the turbulent atmosphere [62]. It may be written as [63]:

$$r_0 = \left( \frac{16.7}{\lambda^2} \int_{path} C_n^2(z) dz \right)^{-\frac{3}{5}} \quad (4)$$

where $C_n^2(z)$ is the (temperature-dependent) atmospheric turbulence strength at the position $z$ along the light path. When the path is a straight line along a zenith angle $\zeta$, the path is longer by a factor approximately equal to $1/\cos(\zeta)$, leading to a smaller Fried parameter:

$$r'_0 = r_0 \left( \cos \vartheta \right)^{3/5} \quad (5)$$

Eq. (4) shows that the Fried parameter increases with wavelength: $r_0 \propto \lambda^{6/5}$. Consequently, an atmospheric turbulence limited telescope will have a seeing that improves slightly with wavelength (i.e., $\theta_{atm}$ becomes smaller; from 808 to 1550 nm the seeing improves by 14%).

Evaluating Eq (1) for two orbital scenarios, one in which the satellite passes directly over the OGS, and one in which it passes at a ground track distance of 500 km, as well as for different values of the Fried parameter, allows us to present in Figure 7 curves of the expected average link attenuation as a function of time. Table 2 summarizes the values of the model parameters used to prepare Figure 7.

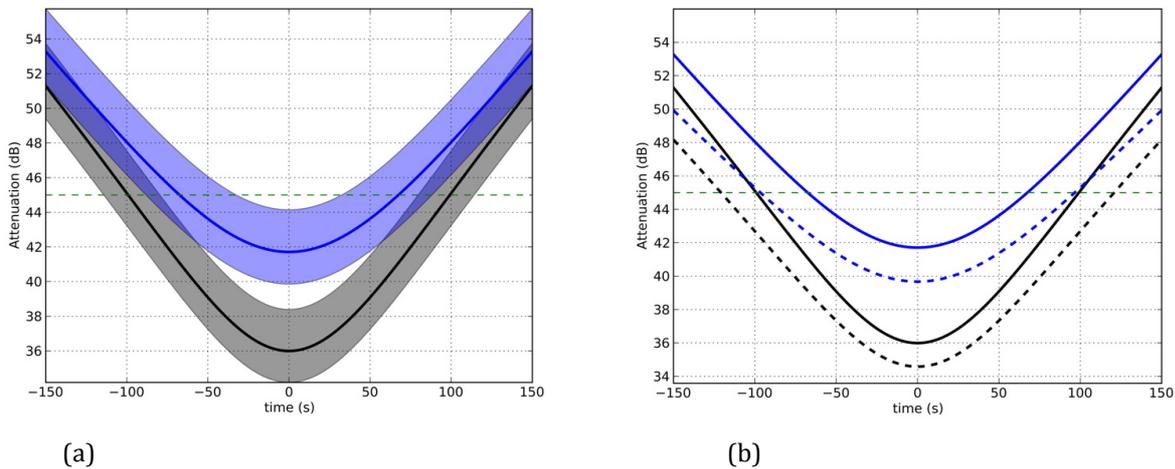

(a)   (b)

Figure 7. **Link losses for two different orbital scenarios**: (a) The lower (black) curve for a passage directly over the OGS and the upper (blue) curve for a ground track that passes at a distance of 500 km from the OGS. The solid curves give the losses for atmospheric turbulence characterized by a Fried parameter $r_0$=20 cm at 808 nm, whereas the shaded bands correspond to the range of 15 cm < $r_0$ < 25 cm. (b) The same as (a) for the solid curves, and in addition the corresponding curves for the link budget at 1550 nm (same atmospheric conditions, corresponding to $r_0$= 44 cm at 1550 nm). The horizontal dashed line indicates the link loss limit of 45 dB for which the experiment duration that QKD would be possible is calculated.

Under conditions of very low atmospheric turbulence ($r_0 >=$ 30 cm at 810 nm, $>=$ 65 cm at 1550 nm), the link attenuation is always smaller than 45 dB during the 240 s of the orbit reserved for the QKD experiment. Figure 6 shows that such favorable conditions can be expected to occur only about 9 days per year at the ESA OGS at the Canary Islands. Accepting stronger atmospheric turbulence ($r_0$ =20 cm at 810 nm, 40 cm at 1550 nm) means that the link attenuation descends below 45 dB for a smaller fraction of the flight time, reducing the time available for QKD to about 200 s and 140 s for the direct overpass and the distant passing, respectively. Such conditions can be expected during about 112 days per year (*cf.* Figure 6).

Table 2. Link attenuation parameters

| | |
|---|---|
| $\lambda$ | 808 nm / 1550 nm |
| $A_{atm,0}$ | 3 dB / 2 dB |
| $D_R$ | 15 cm |
| $D_T$ | 100 cm |
| $T_R$, $T_T$ | 0.8 |
| $L_P$ | 0.2 |
| $h$ | 550 km |

The link budget has direct consequences for the required data storage and transmission bandwidth. The OGS generates roughly $R=10^8$ (entangled) photon pairs per second. Assuming a lower limit of 40 dB average uplink losses (combined geometric and turbulence losses), this means that the satellite receives on average up to $R_E=10^4$ photons per second. These need to all be time tagged with a resolution $\delta t$ better than the width of the coincidence time window $\tau$, itself limited by detector jitter. The (uncompressed) number of bits that need to be stored with each detector event is thus:

$$bits = \log_2\left(\frac{h}{\delta t}\right) + 2 \tag{6}$$

where $h$ is the experiment duration ("horizon") for which a unique time stamp is required, and the final term accounts for the storage of the polarization information (basis and one of two orthogonal directions). The number of bytes is then obtained as *bytes* = $R_E \cdot$(*bits*/8). Taking the rather conservative values of $h$ = 6 months and $\delta t$ = 25 ps, we obtain *bits* = 61.1, or 64 bits after rounding off. The byte rate is then 80 kB/s. For a typical experiment of < 5 minutes duration this requires storage of 24 Mbytes per experiment. With a maximum of 3 passes per day, this comes to 72 MB per day. To this one needs to add house keeping data such as critical temperatures, GPS and star tracker data, etc., that however can be sampled at much lower rate, e.g., just once every second. Even if this would be done continuously throughout the orbital cycles, this would require about 12 MB per day to store 64 values with 2-byte resolution. These numbers are conservative estimates also because in practice the data will be compressed before transmission. E.g., only the first event of each experiment requires a full time stamp, all subsequent events can be stamped relative to the first, saving roughly 16 bits per event, already a 25% reduction in data volume. It is noted that the processing power required to generate the secure key on board of the satellite is not excessive and easily handled by, e.g., a COTS solution incorporating a Zync-based on-board computer (OBC).

## 1.7. KEY RATE

We have performed a study of the expected key rate using a model developed by Ma, Fong, and Lo for QKD with an entangled photon source based on spontaneous parametric down conversion (SPDC) [64]. The model provides an expression for the coincidence detection probability given a source photon (referred to as a "pulse" in the original paper):

$$Q(\mu) = 1 - \frac{1-Y_{0A}}{\left(1+\eta_A \frac{\mu}{2}\right)^2} - \frac{1-Y_{0B}}{\left(1+\eta_B \frac{\mu}{2}\right)^2} + \frac{(1-Y_{0A})(1-Y_{0B})}{\left(1+\eta_A \frac{\mu}{2} + \eta_B \frac{\mu}{2} + \eta_A \eta_B \frac{\mu}{2}\right)^2} \quad (7)$$

Here $\mu$ is the average number of photon pairs produced for one source photon ($\mu < 1$), $\eta_X$ is the detection efficiency of channel $X$ (=A for Alice, or B for Bob), and $Y_{0X}$ is the probability of a dark- or background count in channel $X$ within the coincidence time $\tau$ (s). For a system with $N_{det}$ detectors, a dark count rate of $D_X$ ($X=A,B$), and a background (e.g., due to stray light, poor filtering of beacon light, or other light pollution sources within the FOV of the receiver telescope) count rate of B (s$^{-1}$) in Bob's channel, we can write:

$$Y_{0A} = N_{det} D_A \tau$$
$$Y_{0B} = (N_{det} D_B + B)\tau \quad (8)$$

As in the following we will vary the value of the dark count rate $D_B$, we note here that for the purpose of the simulation, an increase of the dark count rate $D_B$ by and amount $\Delta D$ is equivalent to changing the background count rate B by $N_{det} \cdot \Delta D$ (= $4\Delta D$). The coincidence rate then equals Q times the source photon (singles) production rate (equal to the inverse of the coincidence time window, since the pair production probability is already included in Q):

$$R_{coinc} = \left(\frac{1}{\tau}\right) Q(\mu) \quad (9)$$

We note that the coincidence rate is inversely proportional to the link attenuation until the visibility decreases and the Quantum Bit Error Rate (QBER) increases. This is because dark- and background counts at the NanoBob receiver could accidentally coincide with photon detection at the sender side (Alice, at the OGS), increasing the QBER, and adding to the number of detected coincidences. The rate at which this occurs can be estimated as $N_{acc} = N_t \times N_r \times \tau = (\eta_A R) \times (\eta_B R / A) \times \tau$. Here $N_t$ is the rate of events detected at the sender side, $N_r$ the rate of events detected at the receiver side, $R$ the rate of pair production, and $A$ the link attenuation. For example, with a pair production rate of $R = 10^8$ s$^{-1}$, coincidence time window $\tau = 10^{-9}$ s, and detection efficiency of $\eta = 0.32$, this gives $N_{acc} \approx 10$ cps at a link attenuation of 50 dB, assuming that the sum of dark- and background count rates << $N_r$ = 320 cps. But if the sum of dark and background count rates ($4D_B+B$) is high, say 5000 cps, $N_{acc} \approx 50$ cps (on a total coincidence rate of 63 cps at a link attenuation of 50 dB).

The secret key rate is lower than the coincidence rate since the sequence of coincidences (the "raw key") still contains wrong bits that need to be removed using some kind of error correction. Also, in order to decrease the amount of information that Eve may have been able to obtain, Alice and Bob engage in a process known as privacy amplification that further reduces the number of bits available for the construction of a secret key (see, e.g. [7], [8]). Ma et al. provide a lower limit of the secret key generation ("distillation") efficiency due to post-processing [64]:

$$R_{dist}(QBER) \geq q\big(1 - f(QBER)H_2(QBER) - H_2(QBER)\big) \quad (10)$$

where $q$ represents the basis reconciliation factor, in our protocol equal to 0.5, $f(x)$ is the bidirectional error correction efficiency, and $H_2(x)$ is the binary entropy function: $H_2(x) := -x \log_2(x) - (1-x)\log_2(1-x)$. In the Shannon limit, $f(QBER) = 1$ and the secret key generation fraction reaches zero for QBER → 11.0% [7, 64, 65]. Here, again conservatively, we follow [64] in taking $f(QBER) = 1.22$, in which case the function reaches zero for QBER = 9.4% and secret key distillation is no longer possible. However, the secret key rate is only a factor of 5 lower than the coincidence rate if the QBER ≈ 5%. Only if the QBER exceeds 8%, does the secret key rate drop quickly towards zero.

The QBER could be measured directly in the QKD experiment, but can also be calculated as follows [64]:

$$QBER = e_0 - \frac{1}{Q(\mu)} \frac{(e_0 - e_d)\eta_A \eta_B \mu \left(1 + \frac{\mu}{2}\right)}{\left(1 + \frac{\eta_A \mu}{2}\right)\left(1 + \frac{\eta_B \mu}{2}\right)\left(1 + \frac{\eta_A \mu}{2} + \frac{\eta_B \mu}{2} - \frac{\eta_A \eta_B \mu}{2}\right)} \quad (11)$$

We start our analysis by considering the conservative scenario given by the parameters of Table 3. Notably, we consider that the source produces $10^8$ pairs per second, and that the coincidence time window is limited to 1 ns. This can easily be met by currently existing sources and detection systems that can be integrated on the OGS. We further assume a background count rate of 400 cps. Figure 8 then shows that with a dark count rate of 100 cps per detector, the experiment can tolerate a total link loss up to about 47 dB, and that this limit is reduced to about 40 dB if the dark count rate reaches 1000 cps. The same figure also shows the behavior of the signal-to-noise ratio, defined as SNR = $(N_{max} - N_{min})/N_{min}$, with $N_{min}$ ($N_{max}$) the coincidence count rate measured at the minimum (maximum) of the polarization correlation curve. The SNR may be calculated directly from knowledge of the QBER: SNR = $(1/QBER) - 1$.

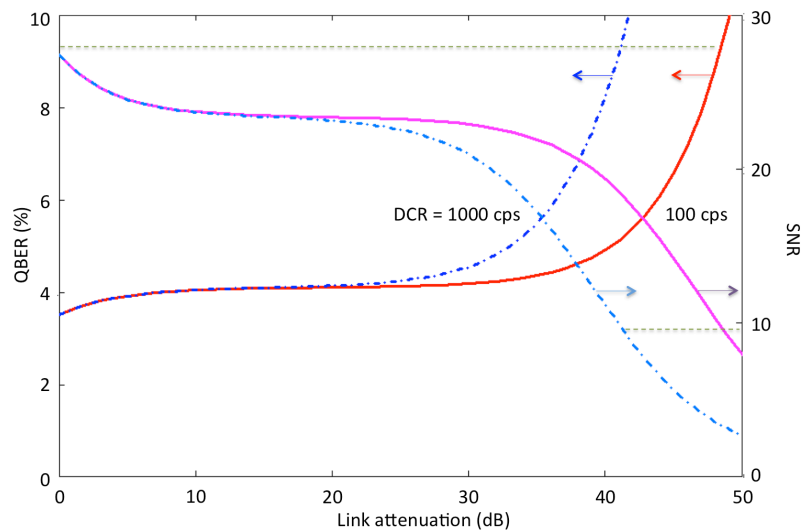

Figure 8. **The calculated QBER and SNR as a function of the link losses** for two different dark count rates (solid red curve: 100 cps; dotted blue curve: 1000 cps per detector). All other parameters are as in Table 3. No secret key distillation is possible if the QBER exceeds 9.4% (SNR > 9.6) for the case that the bidirectional error correction efficiency *f* equals 1.22 (dashed horizontal green line). The corresponding SNR is shown on the right y-axis (solid purple: 100 cps; dashed light blue: 1000 cps dark count rate).

*Table 3. Parameters of the QKD model for a conservative source performance*

| q | basis reconciliation factor | 0.5 | |
|---|---|---|---|
| f(E) | bidirectional error correction function | 1 | |
| τ | coincidence time window | 1 | ns |
| μ | average number of photons per pulse | 0.1 | |
| $D_A$ | OGS dark count rate per detector | 100 | cps |
| $D_B$ | satellite dark count rate per detector | > 100 | cps |
| B | satellite background count rate | 400 | cps |
| $N_{det}$ | number of detectors | 4 | |
| PDE | Photon Detection Efficiency of satellite single photon detectors [40] | 0.4 | |
| $\eta_A$ | OGS overall detection efficiency [44] | 0.6 | |
| $\eta_B$ | $\eta_B = PDE \cdot 10^{-A/10}$, with A the quantum channel link attenuation in dB | | |
| $e_0$ | error probability of dark- and background counts | 0.5 | |
| $e_d$ | error probability of photon arriving on wrong detector (polarization error) | 0.01 | |

The QBER increases and the visibility of the polarization correlation curve (see section 1.4.2) decreases with link attenuation, as well as with increasing dark count rate or background count rate. The visibility may be obtained directly from knowledge of the QBER:

$$V = \frac{1-QBER}{1+QBER} \tag{12}$$

The visibility is a valid estimator of the QBER for the E91 protocol, but not BB84. Using entangled photons, a Bell-test provides a measure of the quantum nature of the link. In order to be able to violate the Bell inequality, the overall visibility *V* should be larger than $1/\sqrt{2} = 0.71$, since the observed Bell parameter (= $V \cdot S_{max}$) should be larger than 2, whereas its quantum mechanical limit $S_{max} = 2\sqrt{2}$. Thus, the SNR should be larger than $2/(\sqrt{2}-1) = 4.83$. As seen above, this condition is always satisfied in the case of a successful QKD experiment.

The visibility for the conditions specified above is shown in figure 9 as a function of link attenuation and for three different levels of dark count rate. As long as the link attenuation does not exceed 51 dB, a dark count rate up to ~250 cps per detector can be accomodated.

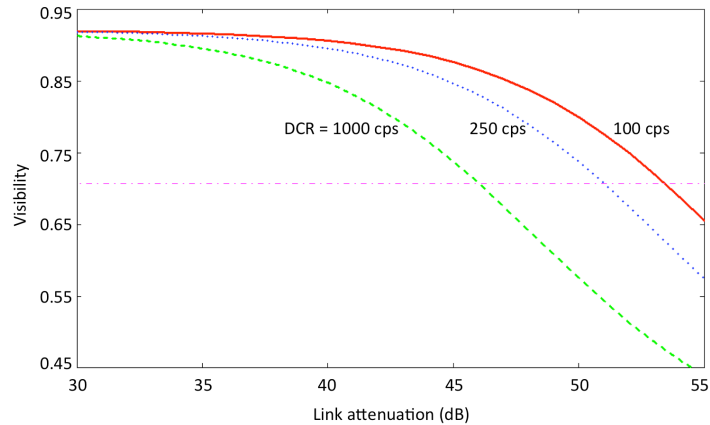

Figure 9. **Visibility as a function of the link attenuation** for three different values of the detector dark count rate (100, 250, and 1000 cps).

A test of the Bell inequalities requires ~1000 coincidences (corresponding to a 3-sigma violation with $S$=2.38 and $\Delta S$=0.126) [7]. With a dark count rate of 100 cps per detector, this can be reached within seconds or less if the link attenuation is less than 40 dB, and within 1 minute if the link attenuation equals ~50 dB, as can be seen by evaluating Eq. (9) with the parameters of Table 3.

In the end, the quantum secured secret key rate is obtained by using Eq. (11) to evaluate the QBER in Eq. (10) as a function of the channel losses and by multiplying the result with the coincidence rate of Eq. (9). The result is shown in Figure 10 (a) for the conservative scenario of Table 3.

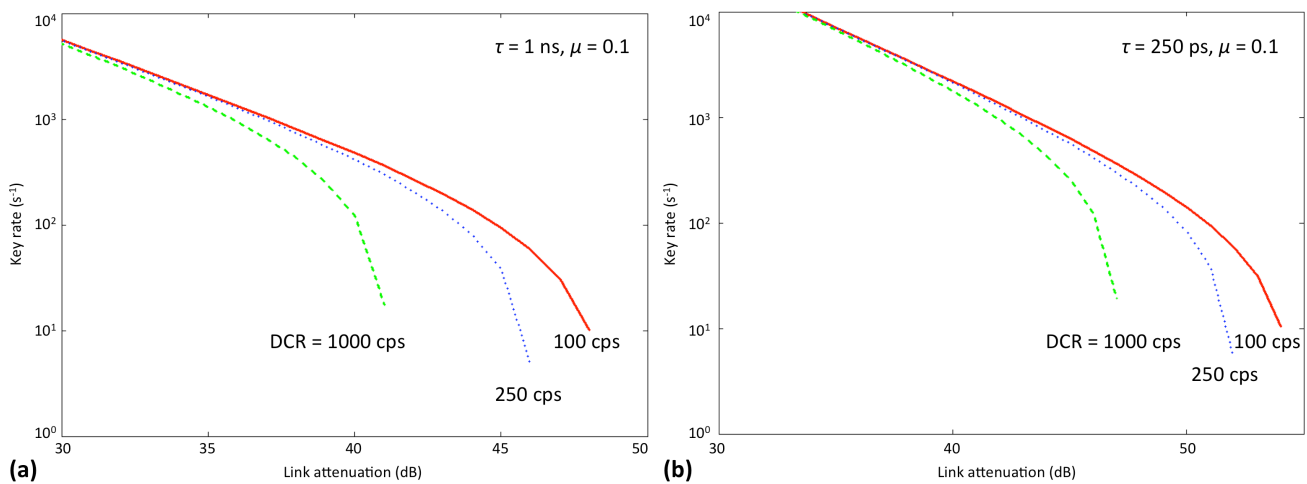

Figure 10. **The secure secret key rate** for three different values of the dark count rate as a function of the link attenuation (solid curve: 100 cps; dotted curve: 250 cps; dashed curve: 1000 cps per detector). (a) For parameters as in Table 3, (b) for the same set of parameters, except for $\tau$ = 250 ps (and thus a pair production of $4 \cdot 10^8$ s$^{-1}$).

The construction of a key of length $10^5$ bits could be accomplished within one ground station overpass (~200 s measurement time) as long as the link attenuation does not exceed 40 dB and the dark count rate is below 250 cps per detector. The Mission Specification of a minimum key length of 1000 bits per experiment (one OGS overpass) can be attained with an average link loss of <45 dB if the dark count rate is lower than about 100 cps. If the detector dark count rates would reach roughly 1000 cps per detector, the maximum link loss that can be sustained is about 38 dB. As we will show further down, this is mostly due to the assumed very conservative coherence time window of 1 ns.

We may now investigate the effect of two important model parameters: the average number of pairs per laser pulse $\mu$ and the coincidence time window $\tau$. Recall that together they determine the pair production rate $R_{\text{pair}} = \mu/\tau$. Increasing $\mu$ while $\tau$ remains constant therefore has the consequence of increasing the pair production rate. This will initially result in a higher key rate, but eventually lead to an accelerated production of accidental coincidences, and an effectively lower key rate. If instead $\mu$ is kept constant and the coincidence time window $\tau$ is reduced to achieve the same increase in pair production rate, the secure key rate increases, and remains at higher levels at high link attenuation. Now a higher pair production rate will enable the experiment to tolerate a significantly higher channel loss.

It appears in fact realistic to expect a coincidence time window shorter than 1 ns. Detectors and electronics should enable reaching 250 ps easily. As mentioned in section 1.4.3 we select single photon detectors with a jitter < 90 ps. The time tagging module itself generally contributes less than 100 ps (see section 1.4.4: the currently persued solution aims for 25 ps maximum and electronic jitter below 10 ps), both on the ground and in the satellite segment. A state-of-the-art OGS polarization analysis module using semi-conducting nanowire single photon detectors could contribute a mere 16 ps to the total time jitter of [44]. Two other effects are expected to lead to only small increases in $\tau$. Two photons that departed the OGS at exactly the same time may still arrive at slightly different times at the satellite, as they may have traversed slightly different path lengths. Beam spreading over the receiver aperture could lead to an increased coincidence time window, but this effect is typically of the order of 1 ps. Also, due to the large velocity at which the satellite moves, uncertainties in its exact position (of the order of tens of cm), will lead to a similar order of magnitude increase in the effective coincidence time window. Together this should lead to a coincidence time window below 200 ps. We therefore have also calculated the expected secure key rate for the case of $\tau$ = 250 ps accompanied by a higher pair production rate of $4 \cdot 10^8$ s$^{-1}$. This is about 30% higher than the value that the Vienna source can currently attain without damage to the SPDC crystal. This can realistically be achieved, e.g., through the implementation of a larger crystal. The result is shown in Figure 10 (b).

From the above analysis we conclude that with conservative parameters for the source performance ($10^8$ pairs/s) and a relatively poor timing resolution ($\tau$ = 1 ns), the experiment can tolerate link losses up to 45 dB by keeping dark and background counts to well below 1000 cps. The secret key rate would reach several times 10 bits/s. Under otherwise the same conditions, but with a source performance as already demonstrated in practice ($3 \cdot 10^8$ s$^{-1}$), and certainly with an improved source as mentioned above, and especially if the coincidence time window can be kept small ($\tau$ < 250 ps), the experiment can accomodate link losses up to 50 dB and still produce a secret key at a rate up to several 100 bits/s. This is shown in Figure 11 for the two orbital scenarios we considered in section 1.6: a direct overpass and a distant overpass in which the ground track passes the OGS at a distance of 500 km under atmospheric conditions characterized by a Fried parameter $r_0$ = 20 cm (at 808 nm). The figure also shows that the length of the secure key (i.e., the integrated key rate) as a function of the time during one OGS encounter.

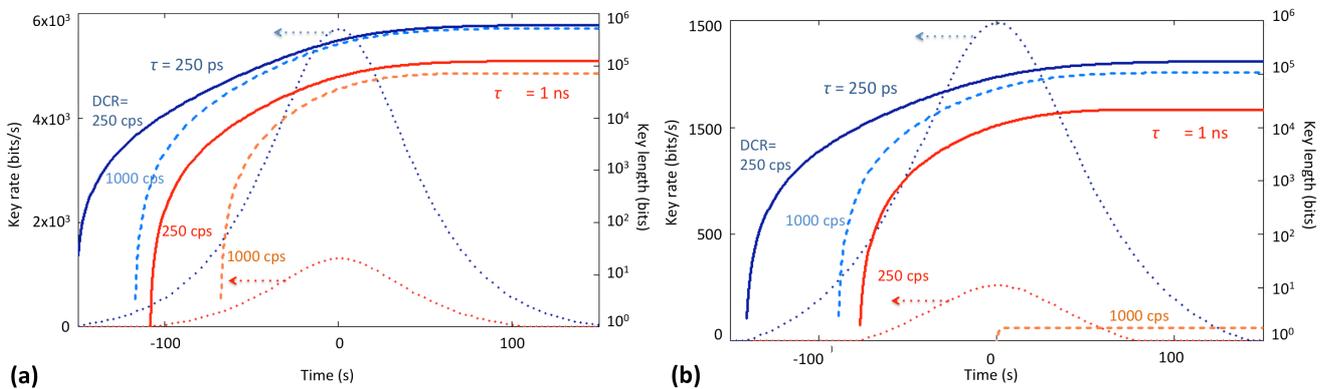

Figure 11. **Secure key production during one overpass** for two scenarios: The lower (red) curves are for the case of a source that generates $10^8$ pairs/s and $\tau = 1$ ns, while the lower (red) curves are for a pair production of $4 \cdot 10^8$ s$^{-1}$ and $\tau = 250$ ps. In both cases the solid curve is for a DCR = 250 cps and the dashed curve for DCR = 1000 cps. The bell curves show the secure key rate (left axis). For reasons of clarity, they are shown only for DCR = 250 cps. The secure key length is given on the right y-axis and for two different values of the dark count rate (250 and 1000 cps). All other parameters are as in Table 3 and the link attenuations are calculated for atmospheric conditions characterized by $r_0 = 20$ cm. The origin of the time axis corresponds to the distance of closest approach (550 km, respectively 743 km). **(a)** for the orbital scenario of a direct overpass, and **(b)** for a overpass at a horizontal distance of 500 km. The key length would be zero bit in case DCR = 1000 cps per detector. The curve shown is therefore for DCR = 945 cps.

## 1.8. DISCUSSION

Cryptography is clearly central to the telecomm industry. Attacks on critical infrastructure components that need to be controlled at a distance, such as satellites, present an obvious concern. Encryption or digital signing of messages using secure keys is one way to fend of such attacks.

Current cryptography standards such as RSA (invented in 1977 by Ron Rivest, Adi Shamir and Len Adleman, [66]) rely on computational complexity and are nowadays the most widely used computer algorithms to encrypt and decrypt messages. With the actual rapid increase of computing power and the increasing likelihood of the arrival of quantum computers in the not-so-distance future, the security offered by RSA, or other schemes using different trap-door mechanisms, will likely decrease rapidly. In fact Peter Schor demonstrated already in 1994 a quantum algorithm able to crack RSA in polynomial time [67, 68]. The eminent arrival of quantum computers clearly poses a serious threat to classical cryptography. As the Chinese Quantum Experiment at Space Scale shows, satellites can make global QKD a reality. However, satellite development has so far been rather complicated and costly. A CubeSat demonstration such as proposed here is therefore not only interesting in its own right and opens up other potential new applications for QKD [68], but also provides important risk-mitigation experience by lowering risk factors for future, larger space missions, potentially aiming for GEO satellite terminals. Spin-offs include atmospheric transmission and turbulence characterization, and dark-area mapping near urban centers, both crucial for future global scale quantum communication efforts.

Miniaturization of CubeSat subsystems, such as those needed for quantum communication, will provide a boost to classical communication technologies and may lead to prototypes for future CubeSat space-qualified subsystems that one day may be available as COTS building blocks for other CubeSat missions.

Although currently not a primary aim, launch of the NanoBob CubeSat in a slightly elliptic orbit will enable the investigation of the gravitational potential on entanglement. The finite speed of light and the description of gravity as space-time curvature are both manifestations of the role of locality in the theory of General Relativity. Quantum theory on the other hand is fundamentally non-local, as manifested by quantum entanglement. These two theories seem difficult to reconcile. (Still, in a controversial paper, it has recently been proposed that entanglement and space-time are linked [70, 71]). Quantum entanglement can be considered to be a linear superposition of two states that is maintained over large distances. General Relativity on the other hand is highly non-linear. The consequences for the interaction of General relativity and quantum theory are currently a hot topic in fundamental physics. Several proposals have appeared in the literature that aim to reconcile the two. A number of papers have suggested that the Schrodinger equation should be replaced by a non-linear equation in the presence of gravity. This would imply that entanglement needs to break down. The proposal by Ralph and Pienaar [72] is particularly attractive and has led the Space-QUEST consortium to propose an entangled photon experiment involving the ISS [4, 22]. In the ISS configuration the theory predicts a significantly different coincidence rate normalized to the single photon rate compared to standard quantum theory. The experiment can in principle also be carried out using NanoBob, provided the satellite is in a slightly elliptic orbit and a sufficiently high photon rate and short coherence time of the source can be achieved. It is estimated that a difference in gravitational field gradient corresponding to an orbital height difference of less than 100 km is needed in order to see an appreciable difference in the decoherence factor for realistic cases of the coherence time (0.8 to 3 ps) [4]. The effect is also predicted to increase with orbital height, making it easier to observe from the 550 km SSO proposed for NanoBob than the ISS orbit at 400 km. Alternatively, the launch of two NanoBob satellites into different circular orbits may still present an economically attractive alternative to the use of an elliptical orbit, given that circular orbits see more and cheaper commercial launch opportunities. It may even be possible to combine data obtained by a single NanoBob satellite with those obtained in a future Space-QUEST experiment on board of the ISS. A limiting factor is likely the required much higher photon rate in order to achieve an adequate signal-to-noise ratio. Increasing the brightness of the source would benefit from larger non-linear crystals, which is already an active area of research. This in turn may require that the photon flux arriving at Alice be distributed over a large number of individual detectors – a costly exercise as it is estimated that roughly a hundred-fold higher photon flux is required. Without reducing the atmospheric losses, or increasing the entanglement efficiency, this implies installing about hundred conventional detector units or using advanced nanowire detectors (about 16 of them) for each polarization direction in the OGS [4]. The space segment is likely not the limiting factor in this experiment. If necessary, the increased data rate could be handled by transferring the data to the ground station during multiple (optical or RF) communication sessions.

## 1.9. CONCLUSION

Our feasibility analysis shows that QKD in an uplink scenario between a ground station and a satellite in LEO is possibly using a space segment that adheres to the 12U CubeSat standard. The SWaP analysis shows that the requirements of volume, weight, and power can be met with a comfortable contingency margin. The design of the receiver telescope with a FOV of 215 μrad guarantees a low background count rate for a ground station located on the Canary Islands (or a similar astronomical observation location) even under the assumption of operation during a full moon phase. At the same time, the FOV is large enough that the required pointing precision is well within reach of current ADCS technology. We have estimated the link budget for an orbital scenario in which the satellite passes directly over the OGS, as well as for one in which its ground track passes at a distance of 500 km. For this we used conservative estimates of the uplink beam spreading due to diffraction and atmospheric turbulence. Subsequently taking conservative parameters for the detection system, and notably a large coincidence time window of 1 ns, we show that the QKD experiment is possible for both orbital scenarios as long as the DCR per detector is not much larger than 250 cps. The secure key length accumulated after one pass would be $1.2 \cdot 10^5$ and $2.1 \cdot 10^4$ bit for the direct and distant overpass, respectively, for a Fried parameter $r_0$ = 20 cm. With a DCR of 1000 cps, the satellite would need to pass almost directly over the OGS to see a reasonable secure key generation rate (that still reaches $7 \cdot 10^4$ bit per pass; however, passing at a horizontal distance of 500 km the secure key length after one pass would be zero bit). This an order of magnitude lower than that reported in an early feasibility study carried out by Rarity and colleagues [34], mostly due to a more conservative and realistic estimate of the atmospheric link losses (an order of magnitude higher: nominally 45 dB versus 35 dB). We have subsequently investigated the effect of increasing the source brightness or decreasing the coincidence time window within still highly realistic limits. Settling on a source pair generation of $4 \cdot 10^8$ s$^{-1}$ and a coincidence time window of 250 ps, both within easy technological reach, we have shown that a secure key rate of the of between $1.7 \cdot 10^5$ and $6 \cdot 10^5$ bits/pass (for, respectively, the distant and the direct overpass, and assuming that up to 300 s of the orbit can be effectively used for QKD) can be reached as long as the DCR of the detectors remain within a factor of ten of their initial DCR (i.e., < 250 cps), also after exposure to radiation in space. There is now growing evidence that Si-APDs, and in especially the thin junction, small diameter types such as we propose to use, will be able to operate in space with such low dark count rates up to one year or longer. Recent reports point towards deep cooling and/or laser annealing as probably successful mitigation strategies [40, 73]. With the shorter coincidence time window a DCR of 1000 cps per detector can be tolerated, yielding calculated secure key lengths of $1.0 \cdot 10^5$ and $5.1 \cdot 10^5$ bits for, respectively, the distant and direct overpasses (with $r_0$ = 20 cm).

Assuming an average key length per pass of $2 \cdot 10^5$ bits and 100 successful passes per year over two selected OGSs, these stations could exchange an absolutely secure key of 20 Mbits per year, or 40 Mbits over the nominal lifetime of 2 years. This is an underestimate, as we have in fact considered that atmospheric conditions with $r_0$ < 20 cm do not contribute at all to total key length and we underestimated to key rate on days that $r_0$ is significantly larger than 20 cm. In fact, a more refined estimate of the maximum key length could be calculated by summing over the contributions of the different bins of the Fried parameter histogram of Fig. 6, and to take into account the exact number of passes and their ground track distances to the OGS for a chosen orbital scenario (although it is of course impossible to know on forehand the exact atmospheric conditions during each OGS encounter; this is, however, an important uncertainty as the distant passes will be more susceptible to poor atmospheric conditions, and the more so the higher the sum of dark and background counts). In any case, counting only the cost of the launch (900 k€), materials and testing costs (600 k€), the direct cost are predicted to be below 40 €/kbit, whereas including labor the cost could still be below 100 €/kbit.

It may be possible to reduce the size of the satellite to 6U or even 3U (see the companion paper in this issue [44]). In the latter case, both volume and power consumption risk becoming the most difficult constraints to satisfy, whereas both the 3U and 6U options entail an obvious penalty of ~6 dB in the link budget due to the two times smaller receiver that can be accommodated. The 12U solution appears for the moment the preferred compromise, considering development time, overall cost, performance, and probability of success. The payload could potentially also be carried by a larger LEO satellite, instead of the 12U CubeSat.

A major advantage of the proposed uplink mission scenario is the versatility of the space segment payload, which will be compatible with a variety of QKD protocols, as well as other mission scenarios. These include fundamental physics experiments testing for entanglement decoherence in a gravitational potential and dark area light pollution mapping.

**ABBERVIATIONS**

| | |
|---|---|
| ADCS | Attitude Determination and Control System |
| APD | Avalanche Photo Diode |
| BB84 | Bennett & Brassard 1984 QKD protocol |
| DCR | Dark Count Rate (expressed in counts per second, cps) |
| E91 | Ekert 1991 Entanglement-based QKD protocol |
| FOV | Field Of View |
| LEO | Low Earth Orbit |
| OGS | Optical Ground Station |
| PDE | Photon Detection Efficiency |
| QBER | Quantum Bit Error Rate (normalized to the channel capacity) |
| SNR | Signal-to-Noise Ratio |
| SSO | Sun Synchronous Orbit |
| SPD | Single Photon Detector |

**Funding**

The CSUG received funding from the Université Grenoble Alpes, the CNRS, as well as from Air Liquide Advanced Technologies, STMicroelectronics, Teledyne-e2v, Sofradir, and Nicomatic, in the form of corporate patronage. The CSUG is supported by the *Fondation UGA*.
FFG Grant Nr. 4927524 / 847964, FFG Grant Nr. 6238191 / 854022, ESA/ESTEC Grant Nr. 4000112591/14/NL/US